\documentclass[preprint,aps,superscriptaddress,nofootinbib,tightenlines]{revtex4}

\usepackage{epsfig}

\def\OMIT#1{{}}

\newcommand{\gsim}{\raisebox{-0.7ex}{$\stackrel{\textstyle >}{\sim}$ }}
\newcommand{\lsim}{\raisebox{-0.7ex}{$\stackrel{\textstyle <}{\sim}$ }}

\begin{document}

\begin{figure}[!t]
\vskip -1.5cm
\leftline{
{\epsfxsize=1.2in \epsfbox{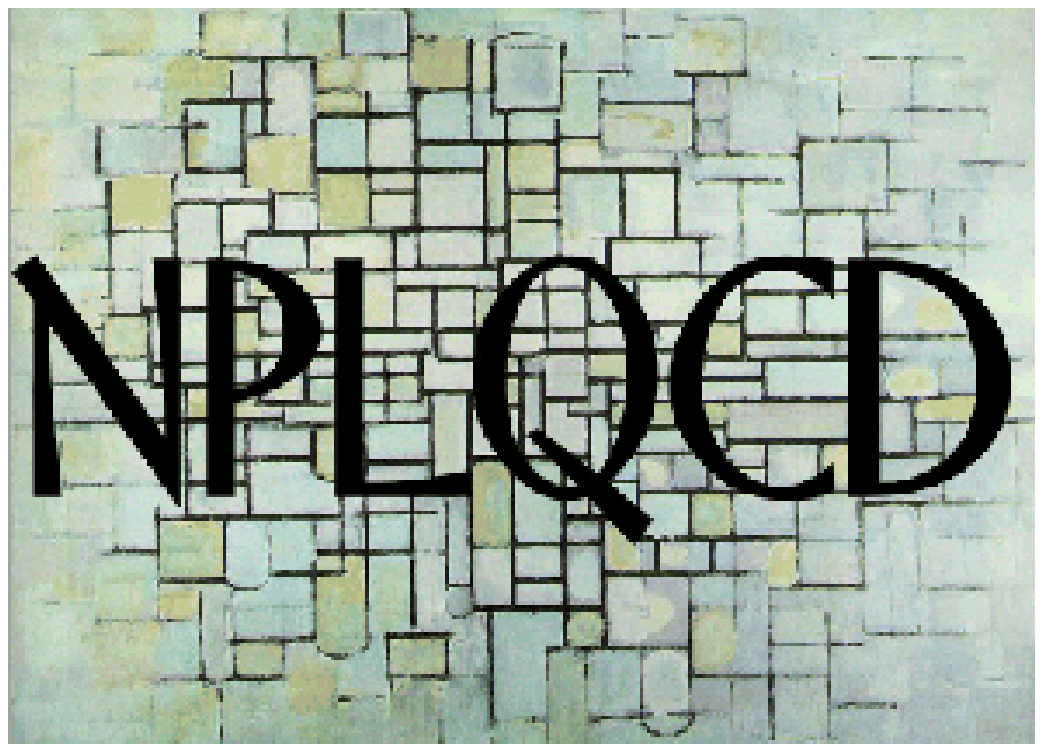}}}
\end{figure}

\preprint{\vbox{
\hbox{UNH-04-04}
\hbox{NT@UW-04-08}
}}

\vphantom{}

\title{Baryon Axial Charge in a Finite Volume}

\author{\bf S.R.~Beane}
\affiliation{Department of Physics, University of New Hampshire,
Durham, NH 03824-3568.}
\affiliation{Jefferson Laboratory, 12000 Jefferson Avenue, 
Newport News, VA 23606.}
\author{\bf M.J.~Savage}
\affiliation{Department of Physics, University of Washington, 
Seattle, WA 98195-1560.\\
\qquad}

\vphantom{}
\vskip 0.5cm
\begin{abstract} 
\vskip 0.5cm
\noindent 

\noindent 
We compute finite-volume corrections to nucleon matrix elements of the
axial-vector current. We show that knowledge of this finite-volume
dependence ---as well as that of the nucleon mass--- obtained using
lattice QCD will allow a clean determination of the chiral-limit values of
the nucleon and $\Delta$-resonance axial-vector couplings.
\end{abstract}

\maketitle

\vfill\eject

%%%%%%%%%%%%%%%%%%%%%%%%%%%%%%%%%%%%%%%%%
\section{Introduction}

\noindent The nucleon axial charge, $g_A$, is a fundamental quantity
in QCD as it in some sense quantifies spontaneous chiral symmetry
breaking in the low-energy hadronic theory. It is known very
accurately from neutron beta decay experiments,
$g_A=1.2670\pm0.0030$ (in units of the vector charge
$g_V$)~\cite{PDG} and therefore serves as an important test of
nonperturbative, first-principles calculations of hadronic properties
using lattice QCD. Fortunately, it is relatively straightforward to
compute $g_A$ in numerical lattice QCD simulations. In spite of this, 
there is still no consensus as regards $g_A$ from lattice
QCD~\cite{Fukugita:1995fh,Liu:1994ab,Gockeler:1996wg,Gusken:1999as,Dolgov:2002zm,Capitani:1999zd,Horsley:2000pz,Sasaki:2003jh}.
In fact, a trend toward under predicting~\cite{Fukugita:1995fh,Liu:1994ab,Gockeler:1996wg} $g_A$ has led to some controversy 
regarding the possibility of large finite-volume effects~\cite{bob,tom}.
A recent quenched simulation using domain-wall fermions over several
volumes finds that large finite-volume effects do seem to account
for the discrepancy~\cite{Sasaki:2003jh}. 
The somewhat tentative current state-of-affairs is primarily due to the fact that current
computational limitations use quark masses, $m_q$, that are
significantly larger than the physical values, lattice spacings, $a$,
that are not significantly smaller than the physical scales of
interest, and lattice sizes, $L$, that are not significantly larger
than the pion Compton wavelength~\cite{Jansen}.  It is confidence in
the extrapolations of these quantities that will allow a confrontation
between lattice QCD predictions for $g_A$ and other hadronic
observables and experiment. While the dependence of $g_A$ on the
lattice parameters can be described by an effective
field theory (EFT), calculability requires maintaining the hierarchy
of mass scales: $|{\bf p}|\; ,\; m_\pi\ \ll \ \Lambda_\chi\ \ll \
a^{-1}$, where $|{\bf p}|$ is a typical momentum in the system of
interest, $m_\pi$ is the pion mass and $\Lambda_\chi\sim 2\sqrt{2}\pi
f$ is the scale of chiral symmetry breaking ($f=132~{\rm MeV}$ is the
pion decay constant). 
Lattice simulations are only now
beginning to achieve the hierarchy of scales necessary to utilize a
perturbative extrapolation.

Here we will be concerned primarily with the finite-volume dependence of
$g_A$. In a spatial box of size $L$, momenta are
quantized such that ${\bf p}=2\pi{\bf n}/L$ with ${\bf n}\in{\bf
Z}$. The EFT momentum hierarchy then requires maintenance of the
additional inequality $f L\gg 1$. This bound ensures that (non-pionic)
hadronic physics 
is completely contained inside the lattice volume. 
In addition, the bound $(m_\pi L)^2(f L)^2\gg 1$ 
ensures that the lattice volume has no effect on
spontaneous chiral symmetry
breaking~\cite{Leutwyler:1987ak,Leutwyler:1992yt}.  These two bounds,
taken together, then imply that 
in order to have a perturbative EFT description
$m_\pi L\gsim 1$.  When
$(m_\pi L)^2(f L)^2\lsim 1$, and therefore $m_\pi L\ll 1$, momentum
zero-modes must be treated 
nonperturbatively~\cite{Leutwyler:1987ak,Leutwyler:1992yt} and one is
in the so-called 
$\epsilon$-regime~\footnote{The chiral-limit considerations of Ref.~\cite{bob,tom} fall
in the $\epsilon$-regime. 
However, to our knowledge no systematic 
finite-volume calculation of baryon properties has 
been done in the $\epsilon$-regime.}.

We will consider the range of pion 
masses~\footnote{The current upper limit of this range has
been estimated recently by one of the authors~\cite{searching}}, $130~{\rm
MeV}\lsim m_\pi \lsim 300~{\rm MeV}$, and
therefore we will take $L\gsim 2~{\rm fm}$, keeping in mind that the
EFT may be reaching the limits of its validity when this bound on $L$
is saturated, particularly when the pions are light. For the
observables considered here, finite-volume effects tend to be small
for $L>4~{\rm fm}$. It is therefore of interest to have control over
the finite-volume dependence of hadronic observables in the range
$2~{\rm fm}<L\leq 4~{\rm fm}$.  Chiral perturbation theory ($\chi$PT),
which provides a systematic description of low-energy QCD near the
chiral limit, is the appropriate EFT to exploit the hierarchy of
scales described above and to describe the dependence of hadronic
observables on
$L$~\cite{Leutwyler:1987ak,Gasser:1987zq,betterways,Luscher}.  Recent
work has investigated the finite-volume dependence in the
meson~\cite{golter1,Golterman:1999hv,davidlin,Lin:2002aj,Colangelo:2002hy,Colangelo:2003hf,Becirevic:2003wk,ArLi,Colangelo:2004xr}
sector and in the
baryon~\cite{AliKhan:2002hz,AliKhan:2003kb,AliKhan:2003rw,Khan:2003cu,Kronfeld:2002pi,Beane:2004tw}
sector.

In this paper we compute the leading finite-volume dependence of the
axial-vector charge of the nucleon in heavy-baryon $\chi$PT
(HB$\chi$PT), including the $\Delta$-resonance as an explicit degree
of freedom. The finite-volume corrections to the axial-vector charge
of the nucleon depend on the $\Delta$-nucleon mass splitting and on
the chiral-limit values of the nucleon, $\Delta$-nucleon and
$\Delta$ axial-vector charges.  Traditionally, the nucleon
and $\Delta$ axial couplings have been estimated using the spin-flavor
$SU(4)$ symmetry of the quark model, and in recent work~\cite{Beane:2002td} the
authors have conjectured the chiral-limit values
of these couplings.  We point out that lattice QCD measurements of
finite-volume effects in the axial-vector charge (and mass) of the
nucleon will provide a clean determination of the nucleon and
$\Delta$-resonance axial-vector couplings.

%%%%%%%%%%%%%%%%%%%%%%%%%%%%%%%%%%%%%%%%%%%%%%%%%%%%%%%%%
\section{The Nucleon Axial Charge in a Finite Volume}
\label{sec:axialcharge}

\begin{figure}[!ht]
\centerline{{\epsfxsize=3.0in \epsfbox{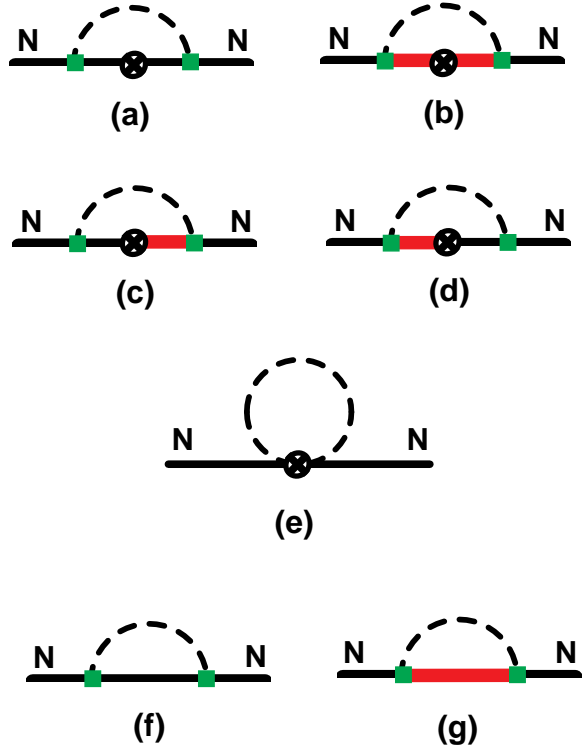}}} 
\vskip 0.15in
\noindent
\caption{\it 
One-loop graphs that contribute
to the matrix elements of 
the axial current in the nucleon.
A solid, thick-solid and dashed line denote a 
nucleon, a $\Delta$-resonance, and a pion, respectively.
The solid-squares denote an axial coupling given in eq.(\ref{eq:intsQCD}),
while the crossed circle denotes an insertion of the axial-vector current operator.
Diagrams (a) to (e) are vertex corrections, while 
diagrams (f) and (g) give rise to wavefunction renormalization.
}
\label{fig:ga}
\vskip .2in
\end{figure}

At one-loop level, the matrix elements of the axial-vector current 
between nucleons of flavor ``a'' and ``b'' may be written as
\begin{eqnarray}
\langle N_b |\ j_{\mu,5}\ | N_a\rangle
& = & 
\left[\ 
\Gamma_{ab}\ +\ c_{ab}
\ \right]
\ 2 \overline{U}_b S_\mu U_a
\ \ \ ,
\label{eq:axmat}
\end{eqnarray}
where $c_{ab}$ represents a counterterm with a single insertion of the 
light-quark mass matrix.
The leading-order Lagrange density describing the interactions between the
pions and the low-lying baryons is
\begin{eqnarray}
{\cal L} & = & 
2 g_A\  \overline{N} S^\mu  A_\mu N
\ +\ 
g_{\Delta N}\ 
\left[\ 
\overline{T}^{abc,\nu}  A^d_{a,\nu}\,  N_b \, \epsilon_{cd} 
\ +\ {\rm h.c.}
\ \right] \ +\ 
2 g_{\Delta\Delta}\  
\overline{T}^\nu S^\mu A_\mu T_\nu  
\ .
\label{eq:intsQCD}
\end{eqnarray}
This Lagrange density gives rise to the diagrams in 
Fig.~\ref{fig:ga}, which are the leading one-loop contributions to the
axial-current matrix elements.
In the isospin limit one finds~\cite{JM}
\begin{eqnarray}
\Gamma_{NN} &=& g_A - i {4\over 3 f^2}
\left[ 4 g_A^3\ J_1(m_\pi,0,\mu)\ +\ 4\left( g_{\Delta N}^2 g_A + {25\over 81}g_{\Delta N}^2
  g_{\Delta\Delta}\right)\ J_1(m_\pi,\Delta,\mu) \
\right.\nonumber\\ &&\left.\qquad\qquad\qquad
+\ {3\over 2} g_A\ R_1(m_\pi,\mu) \  -\  {32\over 9} g_{\Delta N}^2 g_A\  N_1(m_\pi,\Delta,\mu)\ \right]
\label{eq:gainfinite}
\end{eqnarray}
where $J_1(m,\Delta,\mu)$, $R_1(m,\mu)$ and $N_1(m,\Delta,\mu)$ are loop integrals defined in
the Appendix and $\Delta$ is the $\Delta$-nucleon mass splitting. All $\Gamma[\epsilon ]$ poles have been subtracted. They ---and their
associated counterterm $c_{NN}$--- need not concern us here as the finite-volume corrections do not 
depend on the ultraviolet behavior of the theory at leading one-loop order. All of the
couplings (including $f$) in eq.~(\ref{eq:gainfinite}) take their chiral-limit values.

Using the notation
$\delta_L\left(\vartheta\right)\equiv\vartheta(L)-\vartheta(\infty)$ 
to denote the finite-volume corrections to the quantity $\vartheta$,
and the results obtained in the Appendix, 
the finite-volume corrections to $\Gamma_{NN}$ are
\begin{eqnarray}
\delta_L \left(\Gamma_{NN}\right) \equiv \delta g_A
& = &
{m_\pi^2\over 3\pi^2 f^2}\left[\ 
g_A^3 {\bf F_1}
+
\left( g_{\Delta N}^2 g_A + {25\over 81}g_{\Delta N}^2 g_{\Delta\Delta}\right) 
{\bf F_2}
+
g_A  {\bf F_3}
+
g_{\Delta N}^2 g_A {\bf F_4}
\ \right],
\label{eq:gafinitesize}
\end{eqnarray}
where
\begin{eqnarray}
{\bf F_1}(m,L) & = & 
\sum_{{\bf n}\ne {\bf 0}}
\left[\ 
K_0(m L |{\bf n}|) - {K_1(m L |{\bf n}|)\over m L |{\bf n}|}
\ \right];
\nonumber\\
{\bf F_2}(m,\Delta,L) & = & 
-\sum_{{\bf n}\ne {\bf 0}}
\left[\ 
{K_1(m L |{\bf n}|)\over m L |{\bf n}|}
\ +\ {\Delta^2-m^2\over m^2} K_0(m L |{\bf n}|)
\right.\nonumber\\ &&\left.
\qquad -\ 
{\Delta\over m^2}
\int_m^\infty d\beta \ 
{ 2\beta\  K_0(\beta L |{\bf n}|) +(\Delta^2-m^2) L |{\bf n}|\  K_1(\beta L |{\bf n}|)
\over \sqrt{\beta^2+\Delta^2-m^2}}
\right];
\nonumber\\
{\bf F_3}(m,L) & = & 
-{3\over 2} \sum_{{\bf n}\ne {\bf 0}} {K_1(m L |{\bf n}|)\over m L |{\bf n}|};
\nonumber\\
{\bf F_4}(m,\Delta,L) & = & 
{8\over 9} \sum_{{\bf n}\ne {\bf 0}} 
\left[\ 
{K_1(m L |{\bf n}|)\over m L |{\bf n}|}
 - 
{\pi e^{-m L |{\bf n}|}\over 2\Delta L |{\bf n}|}
 -
{\Delta^2-m^2\over m^2\Delta}
\int_m^\infty d\beta \ 
{ \beta\  K_0(\beta L |{\bf n}|)\over \sqrt{\beta^2+\Delta^2-m^2}}
\right],
\label{eq:fsdefined}
\end{eqnarray}
and $K_\alpha(z)$ is a modified Bessel function of the second kind.
The extension of this result to PQQCD, including strong isospin violation,
is straightforward to extract from Ref.~\cite{BSPQ} using the results
of this paper. We do not give an asymptotic expression for $\delta g_A$
as we do not find it useful for $L<10~{\rm fm}$ for the pion masses of interest, 
however, it may be found by taking the appropriate asymptotic limits of
eq.~(\ref{eq:fsdefined}) using technology developed in
Ref.~\cite{ArLi,Beane:2004tw}. One sees in eq.~(\ref{eq:fsdefined}) that, as
$m_\pi L\rightarrow 0$, the ${\bf F_i}$ diverge, signaling the transition
to the $\epsilon$-regime and the necessity of a non-perturbative resummation.

%%%%%%%%%%%%%%%%%%%%%%%%%%%%%%%%%%%%%%%%%%%%%%%%%%%%%%%%%
\section{Extracting Axial Charge from Lattice QCD}
\label{sec:disc}

\begin{figure}[!ht]
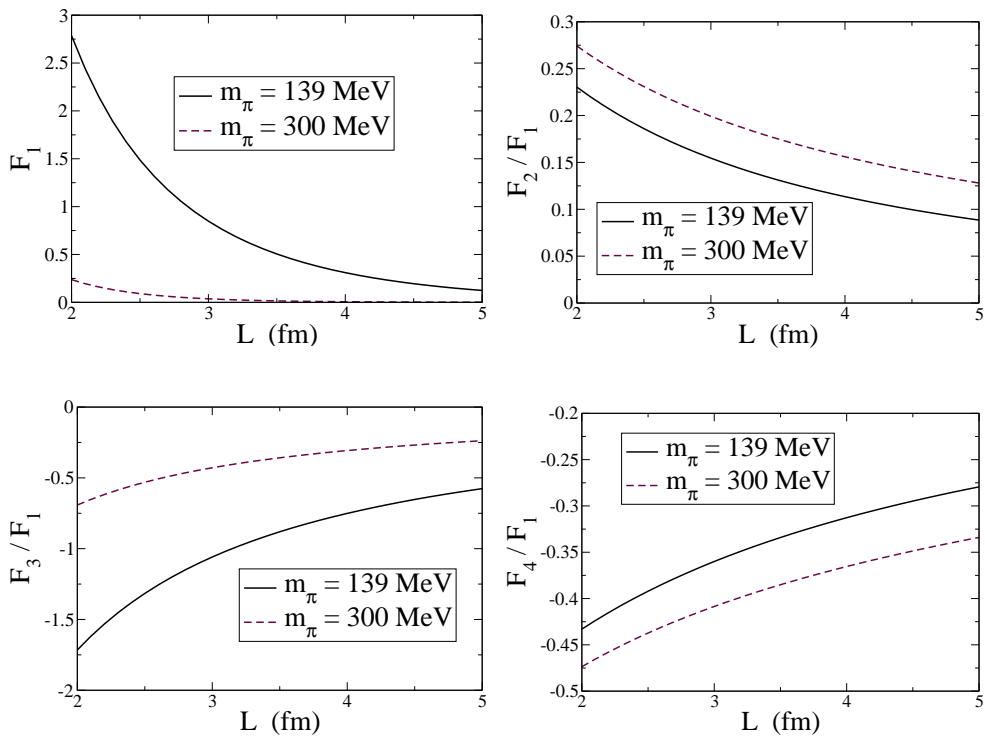

\vskip 0.25in
\centerline{{\epsfxsize=2.5in \epsfbox{F1.eps}}\hskip 0.1in{\epsfxsize=2.5in \epsfbox{F2_F1.eps}}} 
\vskip 0.25in
\centerline{{\epsfxsize=2.5in \epsfbox{F3_F1.eps}}\hskip 0.1in{\epsfxsize=2.5in \epsfbox{F4_F1.eps}}} 
\vskip 0.15in
\noindent
\caption{\it 
Plot of ${\bf F_1}$ and the ratios 
${\bf F_2}/{\bf F_1}$, ${\bf F_3}/{\bf F_1}$ and  ${\bf F_4}/{\bf F_1}$ 
vs. $L$. 
The solid and dashed lines correspond to $m_\pi=139~{\rm MeV}$ and $300~{\rm
  MeV}$, respectively,
for the physical $\Delta$-nucleon mass splitting, $\Delta=293~{\rm MeV}$.}
\label{fig:Fs}
\vskip .2in
\end{figure}

\noindent 
The finite-volume corrections to $\Gamma_{NN}$ depend only on infrared
quantities, {\it i.e.} the axial-vector charges and the pion decay constant, the meson mass and the
$\Delta$-nucleon mass-splitting. Hence, with precise determinations of $f$ (chiral-limit value), $m_\pi$ and $\Delta$, 
lattice data at several different values of $L$ will allow a determination of the axial-vector
charges. However, in order to separate the various contributions to eq.~(\ref{eq:gafinitesize}),
one must ensure that the ${\bf F_i}$ scale differently over the relevant values of $L$.
In Fig.~\ref{fig:Fs} we plot ${\bf F_1}$ and the ratios ${\bf F_2}/{\bf F_1}$, ${\bf F_3}/{\bf F_1}$ and  ${\bf F_4}/{\bf F_1}$ as functions
of $L$ for various pion masses. For ${\bf F_2}$ and ${\bf F_4}$ we use $\Delta=293~{\rm MeV}$.
It is clear from Fig.~\ref{fig:Fs} that the ratios of the ${\bf F_i}$ scale differently and
therefore, in principle, the coefficients of the ${\bf F_i}$ in eq.~(\ref{eq:gafinitesize}) may be extracted
from the $L$ dependence of $\delta g_A$.

\begin{figure}[!ht]
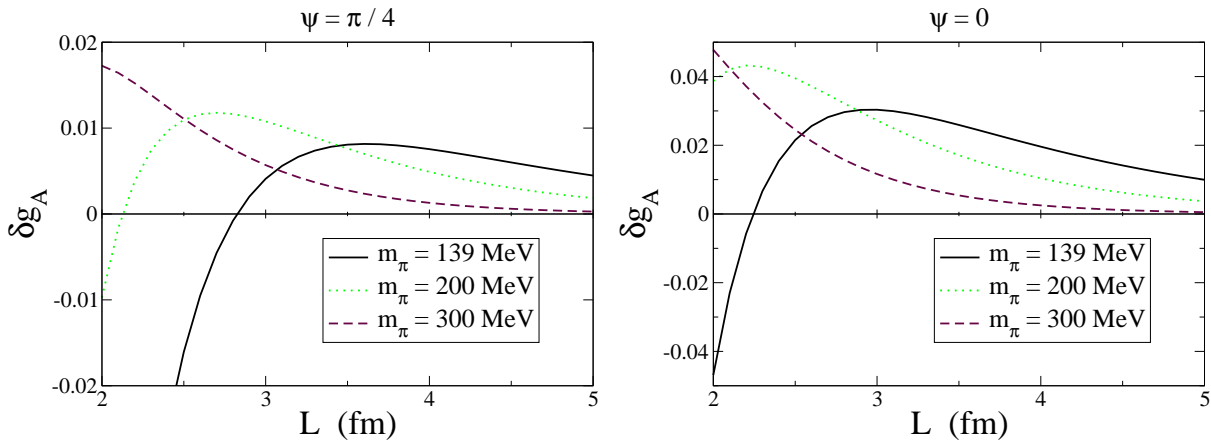

\vskip 0.25in
\centerline{{\epsfxsize=3.1in \epsfbox{delgAvsL.eps}}\hskip0.1in{\epsfxsize=3.1in \epsfbox{delgAvsLsu4.eps}}}
\vskip 0.15in
\noindent
\caption{\it 
The volume dependence of $g_A$ for chiral-multiplet mixing-angles $\psi=\pi/4$
and  $\psi=0$.
The left panel shows $\delta g_A$ vs. $L$ with
$\psi=\pi/4$, where the solid, dotted and dashed lines correspond to $m_\pi=139~{\rm
  MeV}$, $200~{\rm MeV}$ and $300~{\rm MeV}$, respectively. 
The right panel shows $\delta g_A$ vs. $L$ with $\psi=0$ 
(spin-flavor $SU(4)$ values of axial-vector couplings).
The physical $\Delta$-nucleon mass splitting, $\Delta=293~{\rm MeV}$, is used
for both panels.
}
\label{fig:delgA}
\vskip .2in
\end{figure}

In a recent paper by the authors~\cite{Beane:2002td}, based on earlier work by Weinberg~\cite{Weinberga,Weinbergb,Weinbergc,Weinbergd},
it was conjectured that in the chiral limit, the helicity one-half components of the nucleon, $\Delta$ and the Roper ($N'(1440)$) fall into
the reducible $({\bf 2},{\bf 3})\oplus({\bf 1},{\bf 2})$ representation
of $SU(2)_L\otimes SU(2)_R$ with maximal mixing. Denoting the mixing angle
between the irreducible representation as $\psi$ (with maximal mixing
corresponding to $\psi=\pi/4$), the conjecture determines the chiral-limit values
$g_A = 1 + (2/3) \cos^2\psi$, 
$g_{\Delta N}= -2\cos\psi$ and $g_{\Delta\Delta}= -3$.
Inserting these values into eq.~(\ref{eq:gafinitesize}) leads to
\begin{eqnarray}
\delta g_A
&=& {m_\pi^2\over 3\pi^2 f^2}\left[\ 
{\bf F_1} + {\bf F_3} + \left( 2\,{\bf F_1} + \frac{8}{27}\,{\bf F_2} + \frac{2}{3}\,{\bf F_3} + 4\,{\bf F_4} \right)\,\cos^2\psi 
\right.\nonumber\\ &&\left.
+\frac{4}{3}\,\left( {\bf F_1} + 2\, {\bf F_2} + 2\,{\bf F_4} \right) \,\cos^4\psi 
+  \frac{8}{27}\,{\bf F_1}\,\cos^6\psi \ \right] \ .
\label{eq:conjdelgA}
\end{eqnarray}
It would be interesting to have a direct lattice determination of $\psi$ using this formula.
The spin-flavor $SU(4)$ (naive constituent quark-model) 
results are recovered with $\psi=0$. However, the conjectured values
(with $\psi=\pi/4$) are in much better agreement with existing experimental knowledge~\cite{roper,Beane:2002td}.
We use eq.~(\ref{eq:conjdelgA}) to estimate our current knowledge of the finite-volume
dependence of the nucleon axial-vector charge. This expression is plotted as a function 
of $L$ for various pion masses in Fig.~\ref{fig:delgA} for the two cases $\psi=\pi/4$ and $\psi=0$.
Variation of $\psi$ provides a measure of the experimental uncertainty associated with the
chiral-limit values of the axial-vector couplings~\footnote{For a recent discussion of current knowledge
of the chiral-limit value of $g_A$, see Ref.~\cite{searching}.}. It is encouraging that the two scenarios lead
to quite distinct predictions for $\delta g_A$, and therefore a precise
determination of the volume dependence of $g_A$ will allow for a determination
of the mixing-angle $\psi$.
In both cases it is clear that for $L\gsim 2~{\rm fm}$, 
finite-volume effects are at the few-percent level for all relevant pion masses.
%

%%%%%%%%%%%%%%%%%%%%%%%%%%%%%%%%%%%%%%%%%%%%%%%%%%%%%%%%%
\section{Conclusions}
\label{sec:conc}

\noindent 
It has long been known that the infinite-volume S-matrix can
be extracted from power-law suppressed finite-volume effects that arise
when a two-particle 
state is put in a finite 
volume~\cite{yang,Luscher:1986pf,Luscher:1990ux,Beane:2003da,Bedaque:2004kc},
and very recently it has been shown that this method may be extended to include
the effect of external electroweak gauge fields~\cite{Detmold:2004qn}. 
Therefore, a lattice calculation of the energy-levels of a pion and a nucleon
in a finite volume can, in principle, allow for an extraction of the axial-vector couplings.
Important information 
may also be extracted from exponentially-suppressed finite-volume effects that arise from
quantum-loops~\cite{Luscher}. An important observation is that ultraviolet
physics (counterterms) enters the chiral expansion for finite-volume effects beyond leading one-loop
order in the expansion~\cite{AliKhan:2003rw}. Finite-volume effects therefore offer a clean
probe of infrared physics. Moreover, this method is optimal for $m_\pi <\Delta$
where the $\Delta$-resonance is unstable and a direct probe of
$\Delta$ properties is problematic. To conclude, we have computed the leading finite-volume
corrections to nucleon matrix elements of the axial-vector current and shown that
a lattice QCD measurement of this finite-volume dependence 
can determine the chiral-limit values of the
axial-vector charges of the nucleon and $\Delta$-resonance. 

\acknowledgments

\noindent 
The work of SRB is partly supported by DOE contract DE-AC05-84ER40150,
under which the Southeastern Universities Research Association (SURA)
operates the Thomas Jefferson National Accelerator Facility. MJS is
supported in part by the U.S.~Dept. of Energy under Grant
No.~DE-FG03-97ER4014.

%%%%%%%%%%%%%%%%%%%%%%%%%%%%%%%%%%%%%%%%%%%%%%%%%%%%%%%%%
\appendix*{}
\section{Loop Integrals}
\label{sec:app2}

\noindent 
In this appendix we review some standard one-loop integrals 
that arise in HB$\chi$PT~\cite{JM} and
give their finite-volume dependence. First we consider the generic one-loop
integral 
\begin{eqnarray}
I_0(m,\Delta,\mu ) & = & \mu^\epsilon  \int {d^nq\over (2\pi)^n}\ 
{1\over q_0 - \Delta  +  i \epsilon}\ 
{1\over q^2 - m^2 + i \epsilon}
\nonumber\\ 
& = &
{i\over 8\pi^2}
\left[\ 
\Delta\log {{m^2}\over{\mu^2}} - 2 \Delta -
\sqrt{\Delta^2-m^2+ i \epsilon } \ \log\left( { \Delta - \sqrt{\Delta^2-m^2+ i
      \epsilon }\over
 \Delta + \sqrt{\Delta^2-m^2+ i \epsilon }} \right)
\right]
\nonumber\\ 
& = &
-{i\over 8\pi^2} \ \pi {\cal F}(m,\Delta,\mu)
\ \ ,
\end{eqnarray}
where $\pi{\cal F}(m,0,\mu)=\pi m$, $\epsilon=4-n$ and we have subtracted the 
$\Delta \Gamma (\epsilon)$ divergence.
Evaluating the energy integral yields
\begin{eqnarray}
I_0(m,\Delta,\mu ) & = &
 {i\over 2} \mu^\epsilon
\int_m^\infty d\beta\ {\beta\over \sqrt{\beta^2 + \Delta^2-m^2}}\ 
\int {d^{n-1}q\over (2\pi)^{n-1}}\ 
{1\over \left[\ |{\bf q}|^2 + \beta^2\ \right]^{3/2} }
\ \ ,
\end{eqnarray}
where $\beta(\lambda)^2 = \lambda^2 + 2 \lambda\Delta + m^2$ and we have performed
a change of variable that is valid only for $\Delta > 0$, 
as the relation is non-invertible for $\Delta < 0$.
Using the master relation
\begin{eqnarray}
\delta_L\left(  
\int \frac{d^3k}{(2\pi )^3}\ {{1}\over{\left[\ |{\bf k}|^2+{\cal M}^2\ \right]^\alpha}} \right)
& = & 
{{\cal M}^{3-2\alpha}\over 2^{{1\over 2}+\alpha} \pi^{3\over 2}
  \Gamma(\alpha)} 
\sum_{{\bf n}\ne {\bf 0}}  
\left( {\cal M} L |{\bf n}| \right)^{\alpha-{3\over 2}}
K_{{3\over 2}-\alpha} ({\cal M} L |{\bf n}| ),
\label{eq:master}
\end{eqnarray}
which has been derived previously~\cite{Beane:2004tw,AliKhan:2003rw},
one finds the finite-volume corrections 
\begin{eqnarray}
\delta_L\left( I_0(m,\Delta,\mu) \right)
& = &
 {i\over 4\pi^2}
\int_m^\infty d\beta\ {\beta\over \sqrt{\beta^2 + \Delta^2-m^2}}\ 
\sum_{{\bf n}\ne {\bf 0}}
\ K_0 (\beta L |{\bf n}|)
\ \ .
\end{eqnarray}
Notice that there is no renormalization-scale  dependence.
In general, the integral over $\beta$ cannot be performed analytically,
however, for $\Delta=0$
\begin{eqnarray}
\delta_L\left( I_0 (m,0,\mu)
\right)
& = & {i\over 8\pi L} 
\sum_{{\bf n}\ne {\bf 0}}\ {e^{-m L |{\bf n}|}\over |{\bf n}|}
\ \ .
\end{eqnarray}

Next we consider the integral, $I_1(m,\Delta,\mu )$, which appears in the one-loop contribution
to the nucleon mass,
\begin{eqnarray}
I_1(m,\Delta,\mu ) & = & \mu^\epsilon\int {d^nq\over (2\pi)^n}\ 
{(S\cdot q)^2\over v\cdot q - \Delta  +  i \epsilon}\ 
{1\over q^2 - m^2 + i \epsilon}
\nonumber\\ 
& = &
{1\over 4}\left(\ 
\Delta \  R_1(m,\mu)\  +\  \left(\Delta^2-m^2 \right)  I_0(m,\Delta,\mu ) \right)
\ \ ,
\end{eqnarray}
where
\begin{eqnarray}
R_1(m,\mu)\ =\ \mu^\epsilon\int {d^nq\over (2\pi)^n}\ 
{1\over q^2 - m^2 + i \epsilon}\ 
& = & 
{i\over 16\pi^2} m^2 \left[\ 
\Gamma(\epsilon) + 1 - \log {{m^2}\over{\mu^2}}\ \right]
\ \ .
\end{eqnarray}
This integral contributes to the pion-tadpole diagram in Fig.~\ref{fig:ga}(e).
Subtracting the $m^2\Gamma(\epsilon)$ divergence, one then has
\begin{eqnarray}
I_1(m,\Delta,\mu ) & = & 
{i\over 32\pi^2}
\left[\ 
\left( m^2-\Delta^2 \right)
\left( \sqrt{\Delta^2-m^2+ i \epsilon } \ \log\left( { \Delta - \sqrt{\Delta^2-m^2+ i
      \epsilon }\over
 \Delta + \sqrt{\Delta^2-m^2+ i \epsilon }} \right)
\right.\right.\nonumber\\ && \left.\left.\qquad\qquad\qquad\qquad
\ -\ \Delta\log {{m^2}\over{\mu^2}} + 2 \Delta \right)
- {1\over 2} m^2\Delta\log {{m^2}\over{\mu^2}} + {1\over 2} m^2 \Delta
\ \right]
\nonumber\\ 
& = &
{i\over 32\pi^2} F(m,\Delta,\mu)
\ \ .
\end{eqnarray}
Using the master relation, eq.~(\ref{eq:master}), one finds
\begin{eqnarray}
\delta_L \left(\ R_1(m,\mu)\ \right)
\ =\ 
-{i m\over 4\pi^2 L}
\sum_{{\bf n}\ne {\bf 0}} { K_1 (m L |{\bf n}|)\over |{\bf n}|}
\ \ .
\end{eqnarray}
Finally, we find~\cite{Beane:2004tw}
\begin{eqnarray}
\delta_L \left(\ I_1 (m,\Delta,\mu ) \right)
 =  
{i\over 16\pi^2}
\int_m^\infty d\beta {\beta^3\over \sqrt{\beta^2 + \Delta^2-m^2}} 
\sum_{{\bf n}\ne {\bf 0}}
\left[ { K_1 (\beta L |{\bf n}|)\over \beta L |{\bf n}|}  - 
 K_0 (\beta L |{\bf n}|)  \right].
\label{eq:delI1}
\end{eqnarray}

Another useful integral is $J_0(m,\Delta,\mu)={\partial{I_0(m,\Delta,\mu )}/\partial\Delta}$,
\begin{eqnarray}
J_0(m,\Delta,\mu) & = & \mu^\epsilon\int {d^nq\over (2\pi)^n}\ 
{1\over (q_0 - \Delta  +  i \epsilon)^2}\ 
{1\over q^2 - m^2 + i \epsilon}
\nonumber\\ 
& = &
{i\over 8\pi^2}
\left[\ 
\log {{m^2}\over{\mu^2}} - 
{\Delta\over \sqrt{\Delta^2-m^2+ i \epsilon }} \ \log\left( { \Delta - \sqrt{\Delta^2-m^2+ i
      \epsilon }\over
 \Delta + \sqrt{\Delta^2-m^2+ i \epsilon }} \right)
\right].
\end{eqnarray}
The finite-volume corrections are
\begin{eqnarray}
\delta_L\left( J_0 (m,\Delta,\mu)\right) & = & 
-{i L \over 4\pi^2} 
\int_m^\infty d\beta\ \left[ 1 - {\Delta\over \sqrt{\beta^2 + \Delta^2-m^2}}\right] 
\sum_{{\bf n}\ne {\bf 0}}
\ |{\bf n}|\ K_{1} (\beta L |{\bf n}|).
\end{eqnarray}

The one-loop contributions to wavefunction renormalization, Fig.~\ref{fig:ga}(f,g), and to the vertex diagrams, Fig.~\ref{fig:ga}(a,b), for the axial-vector
current operator depend on
\begin{eqnarray}
J_1(m,\Delta,\mu ) & = & \mu^\epsilon\int {d^nq\over (2\pi)^n}\ 
{(S\cdot q)^2\over (v\cdot q - \Delta  +  i \epsilon)^2}\ 
{1\over q^2 - m^2 + i \epsilon}
\nonumber\\ 
& = &
{1\over 4}
\left[\ 
R_1(m,\mu) + 2\Delta I_0(m,\Delta,\mu ) + (\Delta^2-m^2 ) J_0(m,\Delta,\mu)
\ \right]
\nonumber\\ 
& = &
-{3\over 4}\ {i\over 16\pi^2}\ 
\left[\ (m^2-2\Delta^2)\log {{m^2}\over{\mu^2}}
\right.\nonumber\\ && \left.\qquad\qquad\qquad
\ +\ 2\Delta \sqrt{\Delta^2-m^2+ i
      \epsilon }
\ \log\left( { \Delta - \sqrt{\Delta^2-m^2+ i
      \epsilon }\over
 \Delta + \sqrt{\Delta^2-m^2+ i \epsilon }} \right)
\right]
\nonumber\\ 
& = &
-{3\over 4}\ {i\over 16\pi^2}\ 
J(m,\Delta,\mu)
\ \ .
\end{eqnarray}
The finite-volume corrections may be written as
\begin{eqnarray}
\hskip-0.2in\delta_L \left( J_1(m,\Delta,\mu ) \right)= 
-{i\over 16\pi^2}
\int_m^\infty d\beta {\Delta \beta^3\over \left[\beta^2 + \Delta^2-m^2\right]^{3/2}}
\sum_{{\bf n}\ne {\bf 0}}
\left[ { K_1 (\beta L |{\bf n}|)\over \beta L |{\bf n}|}  - 
 K_0 (\beta L |{\bf n}|)  \right] .
\end{eqnarray}

Finally, the vertex diagrams, Fig.~\ref{fig:ga}(c,d), for the axial-vector
current operator depend on
\begin{eqnarray}
N_1(m,\Delta,\mu ) & = & \mu^\epsilon\int {d^nq\over (2\pi)^n}\ 
{(S\cdot q)^2\over v\cdot q - \Delta  +  i \epsilon}\ 
{1\over v\cdot q   +  i \epsilon}\ 
{1\over q^2 - m^2 + i \epsilon}
\nonumber\\ 
& = &
{1\over\Delta}\ \left[\ 
I_1(m,\Delta,\mu ) - I_1(m,0,\mu )\ \right]
\nonumber\\ 
& = &
-{3\over 4}\ {i\over 16\pi^2}\ 
\left[\ (m^2-{2\over 3}\Delta^2)\log {{m^2}\over{\mu^2}}
\right.\nonumber\\ &&\left. \qquad\qquad\qquad
\ +\ {2\over 3}\Delta \sqrt{\Delta^2-m^2+ i
      \epsilon }
\ \log\left( { \Delta - \sqrt{\Delta^2-m^2+ i
      \epsilon }\over
 \Delta + \sqrt{\Delta^2-m^2+ i \epsilon }} \right)
\right.\nonumber\\ &&\left. \qquad\qquad\qquad
+{2\over 3} {m^2\over\Delta}
\left( \pi m - \sqrt{\Delta^2-m^2+ i
      \epsilon }
\ \log\left( { \Delta - \sqrt{\Delta^2-m^2+ i
      \epsilon }\over
 \Delta + \sqrt{\Delta^2-m^2+ i \epsilon }} \right)
\right)
\right]
\nonumber\\ 
& = &
-{3\over 4}\ {i\over 16\pi^2}\ 
K(m,\Delta,\mu)
\ \ .
\end{eqnarray}
The finite-volume corrections are simply
\begin{eqnarray}
\delta_L \left( N_1(m,\Delta,\mu ) \right)
&  = &
{1\over\Delta}\ \left[\ 
\delta_L \left( I_1(m,\Delta,\mu )\right) - \delta_L \left( I_1(m,0,\mu )\right)\
\right]
\ \ ,
\end{eqnarray}
where one uses eq.~(\ref{eq:delI1}).

\vfill\eject

\end{document}